\documentclass[aps,prd,superscriptaddress,12pt,showpacs,notitlepage]{revtex4}
\usepackage{amsfonts}
\usepackage{graphics}
\usepackage{graphicx}

\newcommand{\be}{\begin{equation}}
\newcommand{\ee}{\end{equation}}
\newcommand{\bea}{\begin{eqnarray}}
\newcommand{\eea}{\end{eqnarray}}
\newcommand{\pa}{\partial}
\newcommand{\bb}{\bibitem}
 
\begin{document}
\title{$N=1$ supersymmetric extension of the baby Skyrme model}
\author{C. Adam}
\affiliation{Departamento de F\'isica de Part\'iculas, Universidad de Santiago de Compostela and Instituto Galego de F\'isica de Altas Enerxias (IGFAE) E-15782 Santiago de Compostela, Spain}
\author{J.M. Queiruga}
\affiliation{Departamento de F\'isica de Part\'iculas, Universidad de Santiago de Compostela and Instituto Galego de F\'isica de Altas Enerxias (IGFAE) E-15782 Santiago de Compostela, Spain}
\author{J. Sanchez-Guillen}
\affiliation{Departamento de F\'isica de Part\'iculas, Universidad de Santiago de Compostela and Instituto Galego de F\'isica de Altas Enerxias (IGFAE) E-15782 Santiago de Compostela, Spain}
\author{A. Wereszczynski}
\affiliation{Institute of Physics,  Jagiellonian University,
       \\ Reymonta 4, Krak\'{o}w, Poland}


\begin{abstract}
We construct a method to supersymmetrize higher kinetic terms and apply it to the baby Skyrme model. We find that there exist $N=1$ supersymmetric extensions for  baby Skyrme models with arbitrary potential. 
\end{abstract}

\maketitle 

\section{Introduction}
The interest in topological soliton models has been rising ever since their discovery, both because of their rich intrinsic mathematical structure and due to a large field of possible applications, ranging from particle physics to condensed matter systems. One interesting question concerning topological soliton models is whether they allow for supersymmetric extensions and whether other mathematical properties of (some of) the models, like the existence of Bogomolny bounds and corresponding BPS solutions, may be related to the supersymmetric extensions and their properties, like central extensions in the corresponding SUSY algebra. In 1+1 dimensions, simple scalar field theories consisting of a standard kinetic and potential term support topological solitons if the potential allows for more than one vacuum. Further, it has been known for a long time that these simple models allow for  supersymmetric extensions \cite{divecchia-ferrara}, and that the corresponding SUSY algebra has  a central extension where the central charge is related to the topological charge of the soliton \cite{witten-olive}. In higher dimensions, on the other hand, as a result of the Derrick theorem simple scalar field theories do not support, in principle, topological solitons and, therefore, one has to introduce more structure.     

One possibility consists in the inclusion of gauge fields, and it is well-known that the resulting theories, like the abelian Higgs or the Chern--Simons Higgs models in 2+1 dimensions, the BPS monopole model in 3+1 dimensions, or pure Yang--Mills theory in 4+0 dimensions, allow for supersymmetric extensions and that their topological charges are reflected in the central extensions of the corresponding SUSY algebras
\cite{edelstein-nunez}, \cite{d'adda-horsley}, \cite{d'adda-divecchia}, \cite{witten-olive}.  

Another possibility to circumvent the Derrick theorem in higher dimensions is to allow for non-standard kinetic terms, usually higher (than second) powers of first derivatives in the Lagrangian.  The probably best-known model of this type which allows for topological solitons is the Skyrme model \cite{skyrme} in 3+1 dimensions with the group SU(2) as the field (target) space. Much less is known about supersymmetric extensions of this second type of topological soliton models.
The supersymmetric extensions of a $S^2$ (or CP(1)) restriction of the Skyrme model (the so-called Skyrme--Faddeev--Niemi (SFN) model) were investigated in \cite{nepo} and in \cite{frey}. In both papers, a formulation of the SFN model was used where the CP(1) restriction of the Skyrme model is achieved via a gauging of the third, unwanted degree of freedom. As a result, the SFN model is expressed by two complex scalar fields and an undynamical gauge field, which are then promoted to two chiral superfields and a real vector superfield in the Wess--Zumino gauge, respectively. 
The result of the analysis is that the SFN model as it stands cannot be supersymmetrically extended by these methods. Instead, the supersymmetric extension contains further terms already in the bosonic sector, and also the field equations of the bosonic fields are different.

In a different line of development, more general field theories with a non-standard kinetic term - so-called K theories - have been studied with increasing effort during the last years, beginning with the observation about a decade ago of their possible relevance for the solution of some problems in cosmology (k-inflation
\cite{k-infl} and k-essence \cite{k-ess}). K field theories have found their applications in cosmology \cite{bab-muk-1} - \cite{Dzhu1}, and they introduce some qualitatively new phenomena, like the formation of solitons with compact support, so-called compactons \cite{werle} - \cite{fring}.    Quite recently, investigations of the problem of possible supersymmetric extensions of these K field theories have been resumed \cite{bazeia2}, \cite{susy1}, \cite{ovrut1}, \cite{ovrut2}. Here, \cite{bazeia2} and \cite{susy1} studied supersymmetric extensions of K field theories in 1+1 and in 2+1 dimensions, whereas the investigations of \cite{ovrut1} and \cite{ovrut2} are for 3+1 dimensional K theories, and with some concrete cosmological applications (ghost condensates and Galileons) in mind. 

It is the purpose of this letter to explicitly construct an $N=1$ supersymmetric extension of the baby Skyrme model. The baby Skyrme model is a model supporting topological solitons in 2+1 dimensions, with a $S^2$ target space \cite{piette1}, \cite{piette2}, \cite{weidig}. 
For some recent results see e.g., \cite{bS1}, \cite{speight2}.
Its field contents and its Lagrangian are like the ones of the SFN model, but the topology is more similar to the Skyrme model (solitons are classified by a winding number, not by a linking number like in the SFN model). The baby Skyrme model serves, on the one hand, as a simpler toy model to study general features of topological solitons. Its supersymmetric extensions will, therefore, be interesting for the general understanding of the role of supersymmetry in topological soliton models, as well.  
On the other hand, the baby Skyrme model has found some applications, especially in condensed matter physics, e.g. for the description of quantum Hall ferromagnets \cite{sondhi} or of spin textures \cite{yu-etal}, \cite{ezawa1}. 
The supersymmetric extension method we use is, in fact, similar to the methods used in \cite{ovrut1}, \cite{ovrut2}, but adapted to the case of  2+1 dimensions with its specific spin representation of the Lorentz group and its specific SUSY algebra. We shall find that our supersymmetric extension method may be applied to each term in the baby Skyrmion Lagrangian separately, which explains why it may be applied to arbitrary baby Skyrme models, in principle even allowing for the addition of further terms which do not belong to the standard baby Skyrme models.

\section{Supersymmetric baby Skyrme models}

The class of baby Skyrme models we shall consider in this letter is given by the Lagrangian
\be \label{BS-lag}
L= \frac{\lambda_2}{2} L_2 + \frac{\lambda_4}{4} L_4 + \frac{\tilde \lambda_4}{4} \tilde L_4 + \lambda_0 L_0
\ee
where the $\lambda_i$ are coupling constants and the $L_i$ are (the subindices refer to the number of derivatives)
\be 
L_2 = \partial_\mu \vec \phi \cdot \partial^\mu \vec \phi 
\ee
(the standard nonlinear sigma model term),
\be
L_4 = -(\partial_\mu\vec{\phi}\times\partial_\nu\vec{\phi})^2
\ee
(the Skyrme term),
\be
\tilde L_4 = (\partial_\mu\vec{\phi}\cdot\partial^\mu\vec{\phi})^2
\ee
(another quartic term),
and 
\be
L_0 = -V(\phi_3)
\ee
is a potential term which is usually assumed to depend only on the third component $\phi_3$ of the field. The three-component field vector $\vec \phi$ obeys the constraint $\vec \phi^2 =1$. The term $\tilde L_4$ is absent in the baby Skyrme model (i.e., $\tilde \lambda_4 =0$), but this term is considered in some extensions of the model, especially in the corresponding model in one dimension higher (the SFN model in 3+1 dimensions), see, e.g., \cite{ferr}. Further, we shall see that our supersymmetric extension can be applied to each term separately, therefore we include the $\tilde L_4$ term in the discussion for the sake of generality. 

The field theories we consider exist in 2+1 dimensional Minkowski space, and our supersymmetry
conventions are based on the widely used ones of \cite{Siegel}, where
our only difference with their conventions is our choice of the
Minkowski space metric $\eta_{\mu\nu} = {\rm diag} (+,-,-)$. All sign differences between this paper and \cite{Siegel} can be traced back to this difference.
We introduce three $N=1$ real scalar superfields, i.e.
\begin{equation}
\Phi^i(x,\theta)=\phi^i(x)+\theta^\alpha\psi^i_\alpha (x)-\theta^2F^i(x),\quad i=1,2,3
\end{equation}
where $\phi^i$ are three real scalar fields, $\psi^i_\alpha $ are fermionic two-component Majorana spinors, and $F^i$ are the auxiliary fields. Further, $\theta^\alpha$ are the two Grassmann-valued superspace coordinates,
and $\theta^2 \equiv (1/2)\theta^\alpha \theta_\alpha$. Spinor indices are risen and lowered with the spinor metric $C_{\alpha \beta} = -C^{\alpha\beta}=(\sigma_2)_{\alpha\beta}$, i.e., $\psi^\alpha = C^{\alpha\beta}\psi_\beta$ 
and $\psi_\alpha = \psi^\beta C_{\beta\alpha}$. 

The components of superfields can be extracted with the help of the following projections
\be
\label{comp}
\phi(x)=\Phi(z)|,\quad\,\psi_{\alpha}(x)=D_{\alpha}\Phi(z)|,\quad\,F(x)=D^2\Phi(z)|,
\ee
where the superderivative is
\be
D_\alpha = \frac{\partial}{\partial\theta^\alpha} -i \gamma^\mu{}_\alpha{}^\beta
\theta_\beta \partial_\mu \equiv \frac{\partial}{\partial\theta^\alpha} + i\theta^\beta \partial_{\alpha\beta }  \quad , \qquad
D^2 \equiv \frac{1}{2} D^\alpha D_\alpha 
\ee
and the vertical line $|$ denotes evaluation at $\theta^\alpha =0$.

The problem now consists in finding the supersymmetric extensions ${\cal L}_i$ of all the contributions $L_i$ to the Lagrangian (\ref{BS-lag}). For the non-linear O(3) sigma model term $L_2$ this supersymmetric extension was found long ago in \cite{witten1}, \cite{alvarez}. One simply chooses the standard SUSY kinetic term $ D^2 (-\frac{1}{2}D^\alpha \Phi^i D_\alpha \Phi^i)|$ for the lagrangian ${\cal L}_2$ and imposes the constraint $\vec \phi^2 =1$ on the superfield, i.e., $\vec \Phi^2 =1 $, which in components reads
\begin{eqnarray}
\phi^i \cdot\phi^i &=&1\\
\phi^i \cdot \psi ^i_\alpha&=&0\\
\phi^i \cdot F^i &=&\frac{1}{2}\bar{\psi}^a\psi^a
\end{eqnarray}
or, in the purely bosonic sector with $\psi =0$
\begin{eqnarray}
\phi^i\cdot\phi^i&=&1 \label{bos-constr-1} \\
\phi^i \cdot F^i &=&0. \label{bos-constr-2}
\end{eqnarray}  
It may be checked easily that the constraint $\vec \Phi^2 =1$ is invariant under the $N=1$ SUSY transformations
\be
\delta \phi^i =  \epsilon^\alpha \psi^i_\alpha \; , \quad \delta \psi^i_\alpha = -i\partial_\alpha{}^\beta \epsilon_\beta \phi^i - 
\epsilon_\alpha F^i \; , \quad 
\delta F^i = i  \epsilon^\beta  \pa_\beta{}^\alpha \psi^i_\alpha .
\ee
Remark: the fact that the constraint $\vec \Phi^2 =1$ provides just one real constraint in superspace makes it appear natural to consider just $N=1$ supersymmetry. It turns out, nevertheless, that the supersymmetric O(3) nonlinear sigma model possesses an extended $N=2$ supersymmetry, which is not completely obvious in the $N=1$ SUSY formalism, see \cite{witten1}. In fact, all nonlinear sigma models with a Kaehler target space metric have the $N=2$ supersymmetry \cite{zumino1}.

Our task now is to find the ($N=1$) SUSY extensions of the remaining terms in the Lagrangian. As we are mainly interested in the bosonic sector of the resulting theory we shall set the spinor fields equal to zero, $\psi_\alpha^i =0$, in the following. We remark that all spinorial contributions to the lagrangian we shall consider are at least quadratic in the spinors, therefore it is consistent to study the subsector with $\psi^i_\alpha =0$. The following superfields (we display them for $\psi^i_\alpha =0$) are useful for our considerations,
\bea
(D^\alpha \Phi^i D_\alpha \Phi^j)_{\psi =0} &=& 2 \theta^2 (F^iF^j + \partial^\mu \phi^i \partial_\mu \phi^j )  \label{D-eq-1}\\
(D^\beta D^\alpha \Phi^i D_\beta D_\alpha \Phi^j)_{\psi =0} &=& 2(F^iF^j + \partial^\mu \phi^i \partial_\mu \phi^j )  + \nonumber \\
&& 2\theta^2 (F^i \Box \phi^j + F^j \Box \phi^i - \pa_\mu \phi^i \pa^\mu F^j - \pa_\mu \phi^j \pa^\mu F^i )  
\label{D-eq-2} \\
(D^2\Phi^i D^2 \Phi^j)_{\psi =0} &=& F^i F^j + \theta^2 (F^i \Box \phi ^j+ F^j \Box \phi^i ) . \label{D-eq-3}
\eea
We observe that both the product of Eq. (\ref{D-eq-1}) with Eq. (\ref{D-eq-2}) and the product of Eq. (\ref{D-eq-1}) with Eq. (\ref{D-eq-3}) contain terms of the type $F^2 (\partial \phi)^2$, so by choosing the right linear combination we may cancel these unwanted terms. Concretely, we propose the following supersymmetric Lagrangians (remember that $D^2 \theta^2 =\int d^2\theta \theta^2 =-1$)
\be
({\cal L}_2)_{\psi =0} = -\frac{1}{2}[D^2 (D^\alpha \Phi^i D_\alpha \Phi^i)|]_{\psi =0} =  F^iF^i + \partial^\mu \phi^i \partial_\mu \phi^i  
\ee
\bea
(\tilde {\cal L}_4)_{\psi =0} &=& [D^2 ( D^\alpha \Phi^i D_\alpha \Phi^i) (D^2 \Phi^j D^2 \Phi^j - \frac{1}{4} D^\beta D^\alpha \Phi^j
D_\beta D_\alpha \Phi^j )|]_{\psi =0} \nonumber \\
&=& -(F^i)^2 (F^j)^2 + (\pa_\mu \phi^i )^2 (\pa_\nu \phi^j)^2 
\eea
\bea
( {\cal L}_4)_{\psi =0} &=& -\; \frac{1}{2}\epsilon_{ijk}\epsilon_{i'j'k}[D^2 ( D^\alpha \Phi^i D_\alpha \Phi^{i'} D^2 \Phi^j D^2 \Phi^{j'} +
D^\alpha \Phi^j D_\alpha \Phi^{j'} D^2 \Phi^i D^2 \Phi^{i'}  )|]_{\psi =0} \nonumber \\
&& +\; \frac{1}{8}\epsilon_{ijk}\epsilon_{i'j'k}[D^2 ( D^\alpha \Phi^i D_\alpha \Phi^{i'} D^\gamma D^\beta \Phi^j D_\gamma D_\beta \Phi^{j'} +
\nonumber \\
&& \qquad \qquad \qquad
+\;  D^\alpha \Phi^j D_\alpha \Phi^{j'} D^\gamma D^\beta \Phi^i D_\gamma D_\beta \Phi^{i'}
)|]_{\psi =0} \nonumber \\
&=& \epsilon_{ijk}\epsilon_{i'j'k} (F^i F^{i'} F^j F^{j'} - \pa_\mu \phi^i \pa^\mu \phi^{i'} \pa_\nu \phi^j \pa^\nu \phi^{j'} ) = 
-(\pa_\mu \vec \phi \times \pa_\nu \vec \phi )^2
\eea
and for the potential term, as usual 
\be
({\cal L}_0)_{\psi =0} = [D^2 P(\Phi_3) |]_{\psi =0} = F_3 P' (\phi_3) 
\ee
where $P$ is the prepotential and the prime denotes derivation w.r.t. its argument $\phi_3$.
The resulting bosonic lagrangian is
\bea 
({\cal L})_{\psi =0} &=&  \frac{\lambda_2}{2} [(\vec F)^2 + \pa_\mu \vec \phi \cdot \pa^\mu \vec \phi ] + \frac{\tilde \lambda_4}{4}
[(\pa_\mu \vec \phi \cdot \pa^\mu \vec \phi )^2 - ((\vec F)^2)^2 ] \nonumber \\
&& - \; \frac{\lambda_4}{4} (\pa_\mu \vec \phi \times \pa_\nu \vec \phi )^2 + \lambda_0 F_3 P' + \mu_F (\vec F \cdot \vec \phi )
+ \mu_\phi (\vec \phi^2 -1) 
\eea
where $\mu_F$ and $\mu_\phi$ are Lagrange multipliers enforcing the constraints (\ref{bos-constr-2}) and (\ref{bos-constr-1}).

From now on, we restrict to the standard baby Skyrme SUSY extension with $\tilde \lambda_4 =0$ so that the term $\tilde {\cal L}_4$ is absent.
In this restricted case, the (algebraic) field equation for the field $\vec F$ is 
\be \label{F-eq}
\lambda_2  F^i + \lambda_0 \delta^{i3} P' (\phi_3) + \mu_F \phi^i =0.
\ee
Multiplying by $\vec \phi$ we find for the Lagrange multiplier 
\be
\mu_F = -\lambda_0 \phi_3 P' 
\ee
and for the auxiliary field $\vec F$
\be
F^i = \frac{\lambda_0}{\lambda_2} (\phi_3 \phi^i - \delta^{i3})P'
\ee
and, therefore, for the bosonic Lagrangian
\bea
({\cal L})_{\psi =0} &=& \frac{\lambda_2}{2} [(\vec F)^2 + \pa_\mu \vec \phi \cdot \pa^\mu \vec \phi ] 
 - \frac{\lambda_4}{4} (\pa_\mu \vec \phi \times \pa_\nu \vec \phi )^2 + \lambda_0 F_3 P' 
+ \mu_\phi (\vec \phi^2 -1) \nonumber \\
&=& \frac{\lambda_2}{2}  \pa_\mu \vec \phi \cdot \pa^\mu \vec \phi 
 - \frac{\lambda_4}{4} (\pa_\mu \vec \phi \times \pa_\nu \vec \phi )^2 - \frac{\lambda_0^2}{2\lambda_2}(1-\phi_3^2)P'^2 
+ \mu_\phi (\vec \phi^2 -1) .
\eea
This is exactly the standard (non-supersymmetric) baby Skyrme model with the potential term given by
\be
V(\phi_3) = \frac{\lambda_0}{2\lambda_2} (1-\phi_3^2) P'^2 (\phi_3) .
\ee
Obviously, all positive semi-definite potentials $V(\phi_3)$ may be obtained by an appropriate choice for the prepotential $P(\phi_3)$.

Remark: the relation between prepotential and potential differs slightly (by the additional factor $(1-\phi_3^2)$) from the standard SUSY relation between prepotential and potential, due to the constrained nature of the superfield $\vec \Phi$.   

Remark: The baby Skyrme model has a Bogomolny bound in terms of the topological charge (winding number) of the scalar field $\vec \phi$, but nontrivial solutions in general do not saturate this bound. There exist, however, two limiting cases where nontrivial solutions do saturate a Bogomolny bound and solve the corresponding  first order Bogomolny equations. One might wonder whether these limiting cases allow for the supersymmetric extension discussed in this letter, as well. The first limiting case is the case of the pure O(3) sigma model where both the potential and the quartic (Skyrme) term are absent, and, as discussed above, it is well-known that this case has a supersymmetric extension. 
Concerning the second case, it has been found recently that the model without the quadratic O(3) sigma model term (i.e., $\lambda_2 =0$) originally introduced in \cite{Para-Gis}, has nontrivial Bogomolny solutions and, further, an infinite number of symmetries and conservation laws \cite{bS2}, \cite{speight1}. Given the close relation between Bogomolny solutions and supersymmetry, one might expect that this limiting case should have the supersymmetric extension, too, but this is, in fact, not true. The field equation (\ref{F-eq}) for $\vec F$ for the case $\lambda_2 =0$ reads 
$$ \lambda_0 \delta^{i3} P' (\phi_3) + \mu_F \phi^i =0.$$
It does not contain $\vec F$ at all, so $\vec F$ itself is a Lagrange multiplier in this case. For a nontrivial field configuration $\vec\phi$, the only solution of this equation is $\mu_F =0$ and $\lambda_0=0$, therefore the potential term is absent. We conclude that the model consisting only of the quartic Skyrme term ${\cal L}_4$ does allow for a supersymmetric extension, whereas the model consisting of both the quartic Skyrme term and the potential term ${\cal L}_0$ does not allow for the supersymmetric extension discussed in this letter. 

Remark: we also calculated the full Lagrangian with the spinors included. The contributions from ${\cal L}_2$ and ${\cal L}_0$ are just the standard spinor kinetic term and the Yukawa-type coupling term, respectively. The contribution from the Skyrme term ${\cal L}_4$, on the other hand, is quite long (it consists of 17 more terms) and not particularly illuminating, therefore we do not display it here.

\section{Summary}

We described a method to calculate the supersymmetric extensions of higher kinetic terms (K field theories) and applied it to the baby Skyrme model. We found that the baby Skyrme model has a supersymmetric extension which preserves the form of the original (non-supersymmetric) baby Skyrme Lagrangian in the bosonic sector for arbitrary potential. This possibility to supersymmetrize the baby Skyrme model seems to have gone unnoticed up to now, probably because of some inherent difficulties in the supersymmetrization of higher K terms. Indeed, in general supersymmetric extensions of higher kinetic terms tend to render the "auxiliary" field dynamical, or at least to couple it to field derivatives, which in turn drastically changes the behaviour of the field theory under consideration. Also, higher kinetic terms tend to jeopardize the energy balance between bosonic and fermionic degrees of freedom characteristic for standard SUSY theories. We remark that topological soliton models already at the classical or semiclassical level describe relevant degrees of freedom as low-energy limits of more complete quantum field theories in the ultraviolet. As a consequence, the possibility to directly supersymmetrize these topological soliton models is certainly of interest despite the fact that, at this moment, we are not aware of a direct physical application of the baby Skyrme model where supersymmetry is assumed to play a role. 
In addition, the possibility to construct supersymmetric extensions is an interesting mathematical property of a topological soliton model like the baby Skyrme model and might be useful for a better understanding of its theoretical structure. Interestingly, we found that the limiting case of the baby Skyrme model without the quadratic linear sigma model term (that is, the model consisting of ${\cal L}_4$ and ${\cal L}_0$), does not allow for the supersymmetric extension in spite of its infinitely many exact Bogomolny solutions and its infinitely many symmetries \cite{bS2}. 

A further problem of interest concerns the possibility to apply the supersymmetric extension method presented in this letter to further K field theories. In 1+1 and 2+1 dimensions the supersymmetric extension is rather straight forward   
and may be applied to a quite general class of K field theories. A more detailed discussion of these issues will be published elsewhere. In 3+1 dimensions, on the other hand, the class of K field theories which admit a supersymmetric extension might be more restricted. There, the simplest superfield is the chiral superfield with a {\em complex} scalar field in the bosonic sector, which implies some restrictions on the field contents of theories amenable to supersymmetric extensions. Nevertheless, it might be possible to supersymmetrize topological soliton models in 3+1 dimensions by first choosing a field contents in accordance with the requirements of 3+1 dimensional supersymmetry, and by then introducing the constraints necessary for the reduction of the degrees of freedom to the soliton model one wants to investigate. 
We finally remark that,
as already stated, supersymmetric extensions for some K field theories in 3+1 dimensions with applications in cosmology have been studied recently in \cite{ovrut1}, \cite{ovrut2}, using analogous methods.

\section*{Acknowledgement}
The authors acknowledge financial support from the Ministry of Science and Investigation, Spain (grant FPA2008-01177), 
the Xunta de Galicia (grant INCITE09.296.035PR and
Conselleria de Educacion), the
Spanish Consolider-Ingenio 2010 Programme CPAN (CSD2007-00042), and FEDER.

\end{document}